\documentclass[12pt,epsf]{article}
\usepackage[xdvi]{graphicx}
\newcommand{\beq}{\begin{equation}}
\newcommand{\eeq}{\end{equation}}
\newcommand{\bea}{\begin{eqnarray}}
\newcommand{\eea}{\end{eqnarray}}
\makeatletter

\@addtoreset{equation}{section}
\makeatother

\begin{document}
\setlength{\baselineskip}{18pt}
\begin{titlepage}

\begin{center}
{\large\bf Structure of S and T Parameters   
in Gauge-Higgs Unification}\footnote{Talk given at 2006 International 
Workshop SCGT 06 "Origin of Mass and Strong Coupling Gauge Theories" 
21-24 November 2006, Nagoya, Japan.} 
\end{center}
\vspace{20mm}

\begin{center}
C.~S. Lim\footnote{e-mail: lim@kobe-u.ac.jp}
and Nobuhito Maru\footnote{e-mail: 
maru@people.kobe-u.ac.jp, Speaker. 
This talk is based on \cite{LM}.}
\end{center}
\vspace{1cm}
\centerline{{\it Department of Physics, Kobe University,
Kobe 657-8501, Japan}}
%
%
\vspace{2cm}
\centerline{\bf Abstract}
\vspace{0.5cm}
We investigate the divergence structure of one-loop corrections to 
S and T parameteres in gauge-Higgs unification. 
We show that these parameters are finite in five dimensions, 
but divergent in more than six dimensions. 
Remarkably, a particular linear combination of 
S and T parameters becomes finite in six dimension case, 
which is indicated from the operator analysis in a model independent way. 
\end{titlepage}

\section{Introduction}
Solving the gauge hierarchy problem motivates us to go to 
beyond the Standard Model (SM). 
Gauge-Higgs unification is one of the attractive approach 
to solve the gauge hierarchy problem without supersymmetry. 
In this scenario, Higgs is identified with zero mode of 
the extra component of the gauge field in higher dimensional gauge theories 
and the gauge symmetry breaking occurs dynamically 
through Wilson line phase dynamics. 
One of the remarkable features is that Higgs mass become finite 
thanks to the higher dimensional local gauge invariance. 
Furthermore, 
many applications of gauge-Higgs unification to the real world 
had been carried out in various aspects.

Here we would like to ask the following question; 
Is there any other finite (predictive) physical quantity such as Higgs mass? 
Noting that the gauge-Higgs sector is controlled 
by the higher dimensional local gauge invariance, 
S and T parameters 
\bea
S &=& -16 \pi^2 \Pi'_{{\rm 3Y}}(0), 
\label{Sdef}\\
T &=& \frac{4\pi^2}{M_W^2 \sin^2 \theta_W} 
\left(\Pi_{11}(0) - \Pi_{33}(0) \right), 
\label{Tdef}
\eea
where $\Pi_{ij}(p^2)$ is the $g_{\mu\nu}$ part of the two-point function 
of currents and $\Pi'_{ij} \equiv \frac{d^2}{dp^2}\Pi_{ij}(p^2)$ 
and $\theta_W$ denotes the Weinberg angle, 
are one of the good candidates since these parameters 
are given as the coefficients of dimension six operators composed of 
the gauge fields and Higgs fields. 
In SM, S and T parameters are finite since SM is a renormalizable theory 
and these parameters are coefficients of 
dimension six higher dimensional operators. 
On the other hand, we consider here a nonrenormalizable theory, 
which implies that S and T parameters are in general divergent 
even if they are given by the coefficients of the nonrenormalizable operators. 
However, we know the fact that Higgs mass is finite, which is realized 
thanks to the higher dimensional gauge symmetry. 
Since S and T parameters can be also controlled 
by the higher dimensional gauge symmetry, 
we can expect that these parameters also become finite.

In this talk, we discusss the divergence structure of 
one-loop corrections to S and T parameters 
in the minimal $SU(3)$ gauge-Higgs unification 
on an orbifold $S^1/Z_2$ with a triplet fermion. 
We show that these parameters are finite in 5D case, 
but divergent in more than 6D case. 
The remarkable result is that in 6D case, one-loop corrections to 
S and T parameters themselves are certainly divergent, 
but a particular combination of them becomes finite. 
Its relative ratio agrees with that derived 
from operator analysis in a model independent way. 
This is the crucial difference from 
the universal extra dimension (UED) scenario. 

\section{Model}
We introduce here a minimal model of 5D SU(3) gauge-Higgs unification 
on an orbifold $S^1/Z_2$, whose Lagrangian is given by
\bea
{\cal L} = -\frac{1}{4}F_{MN}F^{MN} + i\bar{\Psi}D\!\!\!\!/ \Psi
\label{lagrangian}
\eea
where $\Gamma^M=(\gamma^\mu, i \gamma^5)$, 
\bea
F_{MN} &=& \partial_M A_N - \partial_N A_M -i g[A_M, A_N]~(M,N = 0,1,2,3,5), \\
D\!\!\!\!/ &=& \Gamma^M (\partial_M -ig A_M), \\
\Psi &=& (\psi_1, \psi_2, \psi_3)^T.
\eea
The periodic boundary conditions for $S^1$ and 
$Z_2$ parities are imposed as follows, 
\bea
A_\mu = 
\left(
\begin{array}{ccc}
(+,+) & (+,+) & (-,-) \\
(+,+) & (+,+) & (-,-) \\
(-,-) & (-,-) & (+,+) 
\end{array}
\right), 
A_5 = 
\left(
\begin{array}{ccc}
(-,-) & (-,-) & (+,+) \\
(-,-) & (-,-) & (+,+) \\
(+,+) &(+,+) & (-,-)
\end{array}
\right), 
\eea
\bea
\Psi = 
\left(
\begin{array}{cc}
\psi_{1L}(+,+) & \psi_{1R}(-, -) \\
\psi_{2L}(+,+) & \psi_{2R}(-, -) \\
\psi_{3L}(-,-) & \psi_{3R}(+, +) \\
\end{array}
\right),
\eea
where $(+,+)$ means that $Z_2$ parities are even at $y=0$ and $y = \pi R$, 
for instance. $y$ is the fifth coordinate and 
$R$ is the compactification radius. 
$\psi_{1L} \equiv \frac{1}{2}(1-\gamma_5)\Psi$, etc. 
One can see that $SU(3)$ is broken to $SU(2) \times U(1)$ 
by these boundary conditions.

Expanding in terms of Kaluza-Klein (K-K) modes and 
integrating out the fifth coordinate, 
we obtain a 4D effective Langrangian for a fermion
\bea
&&{\cal L}_{{\rm fermion}} 
= \sum_{n=1}^{\infty} \left\{  (\bar{\psi}_1^{(n)}, 
\bar{\tilde{\psi}}_2^{(n)}, \bar{\tilde{\psi}}_3^{(n)}) 
\right. \nonumber \\ 
&& \left. \times \left(
\begin{array}{ccc}
i \gamma^{\mu} \partial_{\mu} - m_{n} & 0 & 0 \\
0 & i \gamma^{\mu} \partial_{\mu} -(m_{n} + m + gh) & 0 \\
0& 0 &i \gamma^{\mu} \partial_{\mu} -(m_{n} - m - gh)  
\end{array}
\right)
\left(
\begin{array}{c} 
\psi_1^{(n)} \\
\tilde{\psi}_2^{(n)} \\
\tilde{\psi}_3^{(n)}
\end{array}
\right) \right. \nonumber \\ 
&& \left. 
+\frac{g}{2} (\bar{\psi}_1^{(n)}, \bar{\tilde{\psi}}_2^{(n)}, 
\bar{\tilde{\psi}}_3^{(n)})  
\left(
\begin{array}{ccc}
W_{3}^{\mu}+ \frac{\sqrt{3}B^{\mu}}{3} & W^{+\mu} & W^{+\mu} \\
W^{-\mu} & - \frac{W_{3}^{\mu}}{2} - \frac{\sqrt{3} B^{\mu}}{6} & 
- \frac{W_{3}^{\mu}}{2} + \frac{\sqrt{3} B^{\mu}}{2}  \\
W^{-\mu} & - \frac{W_{3}^{\mu}}{2} + \frac{\sqrt{3} B^{\mu}}{2} & 
- \frac{W_{3}^{\mu}}{2} - \frac{\sqrt{3} B^{\mu}}{6} 
\end{array}
\right) 
\gamma_{\mu} 
\left(
\begin{array}{c}
\psi_1^{(n)} \\
\tilde{\psi}_2^{(n)} \\
\tilde{\psi}_3^{(n)}
\end{array}
\right) \right\} \nonumber  \\ 
&&  + i \bar{t}_{L} \gamma^{\mu} \partial_{\mu} t_{L} 
+  \bar{b} (i \gamma^{\mu} \partial_{\mu} - m - gh) b 
\nonumber \\ 
&& +\frac{g}{\sqrt{2}} (\bar{t}\gamma_{\mu} L b  W^{+\mu} 
+ \bar{b}\gamma_{\mu} L t W^{-\mu}) 
+ \frac{g}{2} (\bar{t}\gamma_{\mu} L t 
- \bar{b}\gamma_{\mu} L b) W_{3}^{\mu} 
\nonumber \\ 
&& + \frac{\sqrt{3}g}{6} (\bar{t}\gamma_{\mu} L t 
+ \bar{b}\gamma_{\mu} L b -2\bar{b}\gamma_{\mu} R b) B^{\mu} 
\label{4Deff}
\eea
where the mass matrix for the non-zero K-K modes 
are diagonalized by use of the mass eigenstates 
$\tilde{\psi}_{2}^{(n)}, \ \tilde{\psi}_{3}^{(n)}$:  
\bea 
\pmatrix{ 
\psi_{1}^{(n)} \cr 
\tilde{\psi}_{2}^{(n)} \cr 
\tilde{\psi}_{3}^{(n)} \cr 
} 
= U 
\pmatrix{ 
\psi_{1}^{(n)} \cr 
\psi_{2}^{(n)} \cr 
\psi_{3}^{(n)} \cr 
}, \ \ \  
U =\frac{1}{\sqrt{2}}
\left(
\begin{array}{ccc}
\sqrt{2} & 0 & 0 \\
0 & 1 & -1 \\
0 & 1 & 1 
\end{array}
\right) 
\eea
and $L \equiv \frac{1}{2}(1-\gamma_5)$, 
$W_\mu^{1,2,3}$, $B_\mu$ are 
the $SU(2), U(1)$ gauge fields, respectively 
and $W_\mu^{\pm} \equiv (W_\mu^1 \pm i W_\mu^2)/\sqrt{2}$. 
$m_{n} = \frac{n}{R}$ is the compactification scale. 
$m = g \langle A_5 \rangle$ is a bottom mass, 
where we consider $\Psi$ to be a third generation quark.  
Dirac particles are constructed as 
\bea 
\psi_{1,2,3}^{(n)} &=& \psi_{1,2,3 R}^{(n)} + \psi_{1,2,3,L}^{(n)}~(n > 0) \\ 
b &=& \psi_{2L}^{(0)}+ \psi_{3R}^{(0)},  
\eea
and the remaining state is a Weyl spinor  
\beq 
t_{L} = \psi_{1L}^{(0)}. 
\eeq
We realized that zero mode part for $t$ and $b$ quarks are exactly 
the same as those in the SM with 
\bea
m_t = 0, \quad m_b = m. 
\label{qmass}
\eea
Thus, we can just use the result in the SM with (\ref{qmass}). 
Note that the mass splitting occurs between the $SU(2)$ doublet component 
and singlet component. 
This pattern of mass splitting has a periodicity with respect to $m$, 
which is a remarkable feature of gauge-Higgs unification.

\section{Calculation of S and T parameters in 5D case}
In this section, we calculate one-loop corrections to T-parameter, 
which is obtained from the mass difference between the neutral W-boson 
and the charged W-bosons 
$\Delta M^2 \equiv \delta \Pi_{33}(0) - \delta \Pi_{11}(0)$. 
The result is given by 
\bea
\Delta M^2 &=& i\frac{3g^2}{16}\frac{2^{D/2}}{D}\sum_{n=-\infty}^\infty 
\int \frac{d^Dk}{(2\pi)^D} \nonumber \\
&\times& \left[
\frac{(2-D)k^2 + D(m_n^2 - m^2)}{[k^2 - (m_n - m)^2][k^2 - (m_n + m)^2]}
-4 \frac{(2-D)k^2 + D(m_n^2 + m^2)}{[k^2 - m_n^2][k^2 - (m_n + m)^2]}
\right]. 
\label{Tall}
\eea
Let us evaluate T-parameter in 5D by carrying out the dimensional 
regularization for 4D momentum integral before taking the mode sum 
in ordet to keep 4D gauge invariance 
and expanding the non-zero mode part of (\ref{Tall}) in $m/m_n$, 
that is, we consider the case where the compactification scale 
is larger than the bottom mass. 
It is straightforward to check that 
the pole terms in $D \to 4$ limit are exactly cancelled 
and the finite value can be calculated from the log terms. 
\bea
\Delta M^2_{(n \ne 0)} = 
-\frac{3g^2}{40\pi^2}\sum_{n=1}^\infty \frac{m^4}{m_n^2} 
= -\frac{g^2}{80}(mR)^2 m^2, 
\label{5DT}
\eea
where $\sum_{n=1}^\infty n^{-2} = \zeta(2) = \pi^2/6$ is used. 
The fact that the leading order term is proportional to $m^4$ 
corresponds to four Higgs vacuum expectation values (VEV) insertions 
in dimension six operator contributing to T-parameter 
$(\phi^\dag D_\mu \phi)(\phi^\dag D^\mu \phi)(\phi: {\rm Higgs~doublet})$. 
The dependence of $m_n^{-2}$ tells us that 
non-zero K-K modes effects are decoupling.

This finite value can be also obtained by taking the mode sum 
before 4D momentum integration. 
If we take the mode sum explicitly in (\ref{Tall}), we find 
\bea
\Delta M^2 &=& -\frac{3g^2}{16}\frac{2^{D/2}}{D}L^{2-D} 
\int_0^1dt \int\frac{d^D\rho}{(2\pi)^D} \nonumber \\
&& \times \left[
-\frac{D}{\rho} \left( \frac{\sinh \rho}{(\cosh \rho -1)} - 1 \right) 
-\frac{D}{2\rho}\left( \frac{\sinh \rho}{(\cosh \rho -\cos \alpha)} - 1 \right) 
\right. \nonumber \\
&& \left. 
-\frac{D}{2} 
\frac{\left(1 + (D-2) \frac{\alpha^2}{\rho^2} \right)}
{\sqrt{\rho^2 + 4t(1-t)\alpha^2}}
\left(
\frac{\sinh \sqrt{\rho^2 + 4t(1-t)\alpha^2}}
{\cosh \sqrt{\rho^2 + 4t(1-t)\alpha^2}-\cos[(2t-1)\alpha]}  -1 \right)
\right. \nonumber \\
&& \left. 
+\frac{D}{2}\frac{\left(4 + (D-2) \frac{\alpha^2}{\rho^2} \right)}
{\sqrt{\rho^2 + t(1-t)\alpha^2}}
\left(
\frac{\sinh \sqrt{\rho^2 + t(1-t)\alpha^2}}
{\cosh \sqrt{\rho^2 + t(1-t)\alpha^2}-\cos [t \alpha]} -1 
\right) \right. \nonumber \\
&& \left. 
- \frac{3D}{2\rho} 
- \frac{\rho^2 + D \alpha^2}{2[\rho^2 + 4t(1-t)\alpha^2]^{3/2}}
+ \frac{4\rho^2 + D\alpha^2}{2[\rho^2 + t(1-t)\alpha^2]^{3/2}}
\right] 
\label{summedT2}
\eea
where 
$L \equiv 2\pi R, \rho \equiv L k$ 
where $k$ is an Euclidean momentum and 
$\alpha \equiv Lm$ (i.e. Aharanov-Bohm phase.).

By performing the dimensional regularization for 4D momentum integral, 
we find 
\bea
\Delta M^2\simeq -\frac{3g^2}{(8\pi)^2}
\left(
m^2 + \frac{4\pi^2}{15}(mR)^2m^2
\right). 
\label{5DTfromsum}
\eea
The $m^2$ is known to be coincide with zero mode contribution. 
The remaining $m^4$ term also agrees with the finite result of 
non-zero K-K mode contributions (\ref{5DT}), which was calculated 
by performing the dimensional regularization 
for 4D momentum before taking the mode sum.

Similarly, one-loop corrections to S-parameter, 
which is obtained from the kinetic mixing term for $U(1)$ gauge bosons, 
can be calculated as 
\bea
\Pi'_{{\rm 3Y}}(0) &=& 
i\frac{\sqrt{3}g^2}{48}2^{D/2}\int \frac{d^Dk}{(2\pi)^D} 
\sum_{n=-\infty}^\infty 
\left[
\frac{2}{(k^2-m_n^2)^2} + \frac{1}{[k^2-(m_n + m)^2]^2} 
\right. \nonumber \\
&& \left. -18\int_0^1 dt t(1-t) 
\left\{
\frac{1}{[k^2 - (m_n + (2t-1)m)^2 - 4t(1-t)m^2]^2} 
\right. \right. \nonumber \\
&& \left. \left. + 2 \frac{(2t-1)[m_n + (2t-1)m]m + 4t(1-t)m^2}
{[k^2 - (m_n + (2t-1)m)^2 -4t(1-t)m^2]^3}
\right\}
\right]. 
\label{allS}
\eea
The finite value is found in a similar way. 
\bea
\Pi'_{3Y}(0) = -\frac{23\sqrt{3}g^2}{120}\frac{1}{(2\pi)^2}
\sum_{n=1}^\infty 
\left(\frac{m}{m_n} \right)^2 
= -\frac{23\sqrt{3}g^2}{2880}(mR)^2. 
\label{finiteS}
\eea
$m^2$ dependence is consistent with the dimension six operator 
representing S-parameter 
$(\phi^\dag W_\mu^a \frac{\tau^a}{2}\phi) B^\mu (\phi^\dag \phi)$. 
$m_n^{-2}$ dependence is also consistent with the decoupling nature 
of K-K particles.

\section{$D>5$ case}
In this section, 
we would like to clarify whether these parameters are finite or not 
in the case higher than five dimensions. 
S and T parameters are given by the coeffcients of 
dimension six operators such as 
$(\phi^\dag W_\mu \phi)B^\mu (\phi^\dag \phi)$ for S-parameter and 
$(\phi^\dag D_\mu \phi)(\phi^\dag D^\mu \phi)$ for T-parameter. 
Naively, the corresponding operators in the gauge-Higgs unification 
can be regarded as the operators where Higgs doublet $\phi$ 
is replaced with $A_i$ ($i$: extra space component index). 
Since $A_i$ transform as $A_i \to A_i + {\rm const}$
by the higher dimensional local gauge symmetry, 
it seems that the local operators for S and T parameters 
are forbidden as in the case of Higgs mass. 
Therefore, we are tend to conclude that S and T parameters 
in gauge-Higgs unification become finite, 
but this argument is too naive, and not correct.

The point is that the gauge invariant local operators for S and T parameters 
are allowed by a single gauge invariant operator 
${\rm Tr}[(D_L F_{MN})(D^L F^{MN})]$. 
Therefore, there is no physical reason for S and T parameters 
to be finite. 
\bea
{\rm Tr}[(D_L F_{MN})(D^L F^{MN})] 
&\supset& 
\frac{1}{2}(8m^4)(W_\mu^3)^2 +(2m^4)W_\mu^+ W^{-\mu} 
+2\sqrt{3} m^2p^2g_{\mu\nu}W^{3\mu}B^\nu
\nonumber \\
&&+2\sqrt{3}m^2(p^2g_{\mu\nu} - p_\mu p_\nu)W^{3\mu}B^\nu. 
\eea
What a remarkable thing is that we can predict some combination of 
S and T parameters although these parameters themselves are divergent. 
We can read off the ratio of them as
\bea
C{\rm Tr}[(D_LF_{MN})(D^LF^{MN})] \to 
\left\{
\begin{array}{l}
\Delta M^2 = 6C m^4 \\
\Pi'_{3Y} = 4\sqrt{3}C m^2, 
\end{array}
\right. 
\eea
where $C$ is an undetermined overall constant. 
Thus, we can expect the combination 
$\Pi'_{3Y} -\frac{2\sqrt{3}}{3m^2}\Delta M^2$ to be finite 
even in more than five dimensions.

In fact, we can show that 
$\Pi'_{{\rm 3Y}}-\frac{2\sqrt{3}}{3m^2}\Delta M^2$ 
is finite in 6D case because 
\bea
\Delta M^2 &=& 
\frac{3g^2}{40\sqrt{2}\pi^2}
\sum_{n=1}^\infty 
\left[
-\frac{m^4}{m_n} +\frac{1}{12}\frac{m^6}{m_n^3}
\right], 
\label{6DT} \\
\Pi'_{{\rm 3Y}}(0) 
&=& \frac{\sqrt{3}g^2}{20\sqrt{2}\pi^2}
\sum_{n=1}^\infty 
\left[ -\frac{m^2}{m_n} +\frac{3}{14}\frac{m^4}{m_n^3}
\right] 
\label{6DS}
\eea
where the first term indicates logatithmic divergence. 
Combining these results (\ref{6DT}) and (\ref{6DS}), 
we obtain the finite result 
\bea
\Pi'_{3Y} - \frac{2\sqrt{3}}{3m^2}\Delta M^2 
= 
\frac{11\sqrt{6}g^2}{3360\pi^2}
m^4 R^3 \zeta(3). 
\label{6Dpdtn}
\eea

\section{Conclusions}
In this talk, we have discussed the divergence structure of 
one-loop corrections to S and T parameters in gauge-Higgs unification. 
Taking a minimal $SU(3)$ gauge-Higgs model 
with a triplet fermion, we have calculated S and T parameters 
at one-loop order. 
In five dimensions, we have shown that one-loop corrections to 
S and T parameters are finite 
and evaluated their finite values explicitly. 
In more than six dimensions, 
S and T parameters are divergent as in the UED scenario. 
However, a particular combination of S and T parameters 
is shown to be finite in six dimension case, 
whose relative ratio was found to agree with that 
derived from the operator analysis in a model independent way. 
This is the crucial difference from the UED scenario.

\subsection*{Acknowledgments}
The speaker (N.M.) would like to thank the organizers for providing me 
an opportunity to present this talk in the conference. 
The work of the authors was supported 
in part by the Grant-in-Aid for Scientific Research 
of the Ministry of Education, Science and Culture, No.18204024.


\end{document}